# Degeneracy removal of spin bands in antiferromagnets with non-interconvertible spin motif pair

Lin-Ding Yuan and Alex Zunger

Renewable and Sustainable Energy Institute, University of Colorado, Boulder 80309, USA

Energy bands in antiferromagnets are generally spin degenerate in the absence of spin-orbit coupling (SOC). Recent studies [Physical Review B **102**, 014422 (2020)] identified formal symmetry conditions for crystals for which this degeneracy can be lifted even in the zero SOC limit. Such materials could enable 'spin-split' antiferromagnetic spintronics without the burden of use of heavy atom compounds. Here, we show that these formal symmetry conditions can be interpreted in terms of easy-to-visualize local motif pair, such as octahedra or tetrahedra, each carrying opposite magnetic moments. Collinear antiferromagnets with such spin motif pair whose components interconvert by neither translation nor spatial inversion will show splitting of spin bands. Such real-space motif-based approach enables an easy way to identify and design of materials having spin splitting without the need for spin orbit coupling, and offers insights on the magnitude of spin splitting.

# I. Introduction

The collusion between physical interactions and the systems' symmetry consists of the law of nature. For example, the relativistic spin orbit coupling (SOC) interaction term in the presence of inversion asymmetry in nonmagnetic crystals results in splitting of spin polarized energy bands (henceforth "spin spitting") known as Rashba [1] and Dresselhaus [2] effects. Likewise, the interaction between electron spin and inhomogeneous magnetization in antiferromagnets was anticipated in 1964 by Pekar and Rashba [3] to results in an unusual spin splitting *independent* of SOC. Yet, many common antiferromagnetic crystals, such as NiO, MnO, do not show such spin splitting. This begs the question of what are the enabling symmetries associated with such a mechanism. Indeed, the obvious need to validate experimentally novel proposed mechanisms calls for establishing the symmetry conditions needed for identifying candidate real materials that could harbor such effects.

The present authors formulated recently the enabling symmetry conditions for the SOC-independent spin splitting.[4,5] Such symmetry conditions disentangle the SOC-independent splitting from the SOC-induced splitting by considering the symmetry at the zero SOC limit [6-8], where spin and space are fully decoupled. This involved utilizing first a few individual symmetry operations: $U$ being a spin rotation of the SU(2) group acting on the spin 1/2 space that reverses the spin; $T$ being spatial translation; $\Theta$ being time reversal, and $I$ being the spatial inversion. These individual operations are then used for constructing two symmetry products: a SOC-free *magnetic symmetry $\Theta IT$*, and a *spin symmetry $UT$* (where the former product can be simplified to $\Theta I$ by proper choice of inversion center). SOC-independent spin splitting [4,5] would occur only when both symmetry products are simultaneously violated. The two symmetry products were then mapped into magnetic space group symmetries [9-14] (that in the most general form include time reversal and consider SOC), thereby allowing the use of the tabulated magnetic structure symmetry information provided in material database [15] to sort out candidate materials [5]. This mapping is further discussed in Appendix A. [16]

Despite being rigorous, the symmetry theory [4,5] did not amount to a transparent structural chemistry intuition for the enabling principles. Alternative model Hamiltonian based approach also fails to provide a straightforward way to identify materials that harbor the spin splitting without SOC. [17-19] The difficulty for a crystal chemist to look at the rule of simultaneous violation of the two symmetry products and suggest predictively a specific real compound following this violation would appear to impede experimental developments in this field. Indeed, experimental efforts in this field are only known for very limited systems, such as $RuO_2$ [20-22].

Here, we transform the formal symmetry filters to simple rules interpreted in terms of classic crystallographic structural motifs [23,24]. These are known in crystal chemistry and include objects such as tetrahedra in zinc-blende crystals, octahedra in perovskite crystals. Here, two generalizations are needed: considering at least two structural motifs and labeling each component by spin. In collinear antiferromagnets, the structural motif entangled with up and down magnetic ordering can be generalized to *spin motif pair* – a pair of structural motifs centered on atoms that carries opposite magnetic moment. Since spin rotation $U$ or the time reversal $\Theta$ act on a collinear antiferromagnet results in a reversed magnetic ordering; The preserving or violation the $UT$ and the $\Theta I$ symmetry products can be interpreted in terms of how the components of the spin motif pair are geometrically related – that is, whether they can or cannot be interconverted by a translation or a spatial inversion (illustrated in Fig. 1). Calculated band structures validate these motif-based rules. Besides, such motif-based description offers extra insights on the magnitude, and momentum-dependence of the spin splitting.

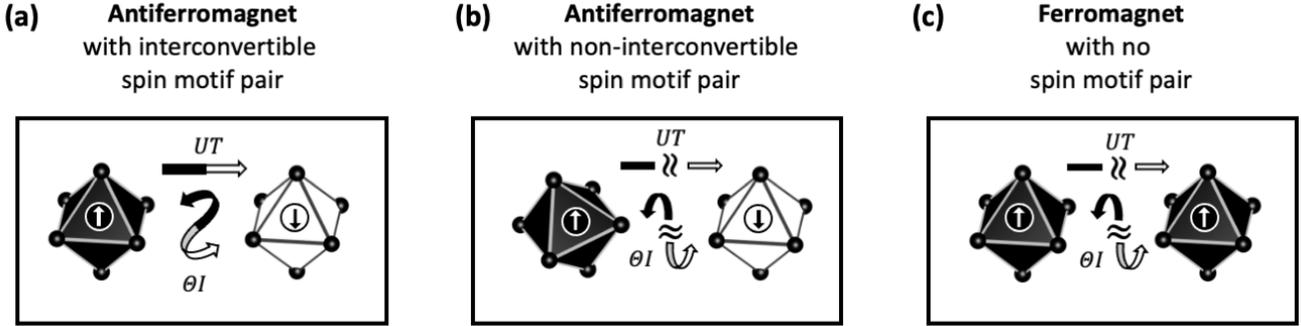

**Figure 1. Magnetic prototypes with motif pair that interconverts or fails to interconvert by inversion and/or translation.** (a) antiferromagnets with spin motif pair interconverts by inversion and/or translation; (b) antiferromagnets with spin motif pair cannot interconvert by inversion nor translation; (c) ferromagnets or ferrimagnets, which has no spin motif pair therefore cannot interconvert by symmetry. The black and white octahedra are used to represent the spin motif pair; The orientation of the magnetic moments is indicated by circled arrows within the shape. The geometric interconvertible relation between spin motif pair is indicated by line arrow (translation related) and S-shaped arrows (inversion related) between them. Broken arrows mean the interconvertible relation does not exist.

## II. Motifs as a descriptor of symmetry

Considering the simplest case where the collinear antiferromagnet has only a single pair of such spin motif in the magnetic unit cell. The preserving (violation) of $UT$ symmetry in a collinear antiferromagnetic compound corresponds to the spin motif pair can (cannot) be interconverted by a non-primitive translation $T$ of the magnetic unit cell. Similarly, the preserving (violation) of $\Theta I$ symmetry corresponds to the two spin motifs can (cannot) be interconverted by a spatial inversion.

Figure 1 illustrate the correspondence in three collinear magnetic prototypes: (1) antiferromagnets with spin motif pair interconverts by inversion and/or translation (shown in Fig. 1(a)). These are antiferromagnets that preserves $\Theta I$ and/or $UT$ symmetry, therefore will not show spin splitting without SOC; (2) antiferromagnets with spin motif pair that interconverts by neither inversion nor translation (shown in Fig. 1(b)). These are antiferromagnets that violated $\Theta I$ and $UT$ symmetry, therefore will show spin splitting even without SOC; (3) Ferromagnets or Ferrimagnets (shown in Fig. 1(c)), which has no paired spin motif. This prototype violates both $\Theta I$ and $UT$ symmetry, therefore will show spin splitting without SOC. Notice there is no restriction on the spin motifs being centrosymmetric or non-centrosymmetric in distinguishing the three prototypes, unless we want to further divide prototype (1) based on symmetry (discussed in Appendix B).

The same correspondence applies for collinear antiferromagnetic compounds with multiple pairs of motifs in the magnetic unit cell. In these cases, the spin motifs of the same magnetic moment shall be grouped; And the interconvertible relation shall be examined for a pair of such spin motif group carrying opposite magnetic moment. For simplicity, we will also refer the "spin motif group pair" as "spin motif pair". As such, collinear antiferromagnetic compounds with spin motif pair convertible by translation or inversion will not exhibit spin splitting without SOC; collinear antiferromagnetic compounds with non-interconvertible spin motif pair will expected to show the unconventional SOC-independent antiferromagnetic induced spin splitting.

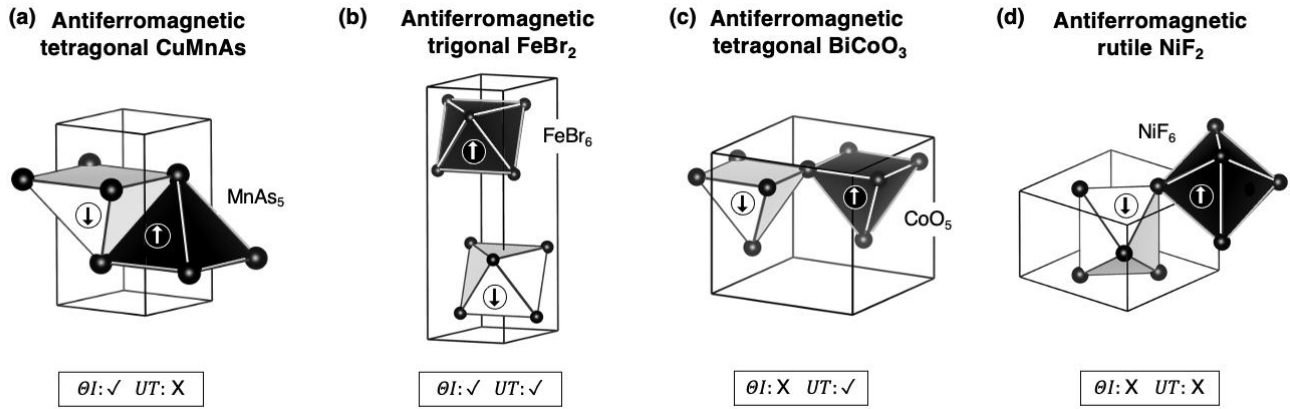

**Figure 2 | Examples of collinear antiferromagnetic compounds with interconvertible or non-interconvertible spin motif pair.** (**a**) Tetragonal CuMnAs with spin motif pair interconverts by inversion; (**b**) Trigonal $FeBr_2$ with spin motif pair interconverts by inversion and translation; (**c**) Tetragonal $BiCoO_3$ spin motif pair interconverts by translation; (**d**) Rutile $NiF_2$ with spin motif pair that fails to interconnect by inversion and translation. Black and white colors are used to map the spin motif pair with opposite magnetic moments. Periodically repeated spin motifs within the same magnetic unit cell are removed to simplify the illustration.

The connection between "non-interconvertible" spin motif pair and the breaking of the effect enabling symmetries readily provides a convenient guide for selecting candidate real antiferromagnetic materials. We will illustrate this in four real antiferromagnetic materials. Figure 2(a) shows the crystal structure of antiferromagnetic tetragonal CuMnAs [25], in which the non-centrosymmetric $MnAs_5$ square-pyramid are alternatively aligned along the [001] direction The two neighboring $MnAs_5$ square pyramid with opposite local magnetic moments constitute the spin motif pair whose components interconverts by inversion but not translation. Similarly, antiferromagnetic trigonal $FeBr_2$ [26] and hexagonal $BiCoO_3$ [27], as shown in Fig. 2(b) and 2(c), respectively, contain spin motif pair whose components interconvert by inversion and/or translation. These three antiferromagnets corresponds to prototype (1) in Fig. 1 and will not show spin splitting. Interestingly, for $BiCoO_3$, the spin splitting is expected to be zero only when SOC vanishes. Theoretical studies [28] show finite spin splitting in this compound when SOC is considered – a case of SOC induced splitting. In contrast to CuMnAs, $FeBr_2$, and $BiCoO_3$, antiferromagnetic rutile $NiF_2$ [29], shown in Fig. 2(d), contains a pair of tetragonal elongated $NiF_6$ spin motifs that is interconverted by neither translation nor inversion (but instead by a 90 degree in-plane rotation). The non-interconvertible geometry of this compound then allows the SOC-independent spin splitting to exist. *Density Functional Theory calculated band structures, provided in Appendix C, agrees with the above motif-based description.*

## III. Motifs beyond being a descriptor of the symmetry

Besides being a descriptor of the symmetry rules, the motif-based approach also offers insights into the physics underpinning: where the spin symmetry breaking originates from, how large is the symmetry breaking. The symmetry breaking information -- entailed in the non-interconvertible spin motifs -- then implies the overall magnitude of the resulting spin splitting effect. We will illustrate these points in an example compound -- lanthanoid perovskite manganite $LaMnO_3$.

The $LaMnO_3$ crystal appears in the perovskites structure (space group: Pnma) consists of corner-sharing $MnO_6$ octahedra motifs. Below the Néel temperature, $LaMnO_3$ is antiferromagnetically ordered which contains oppositely aligned ferromagnetic layers stacked along the c axis (Fig 3). The two ferromagnetic layers (black and white layer in Fig. 3) forms a pair of spin motif. The symmetry-breaking arises from the $a^-b^+a^-$ (Glazer's notation [30]) cooperative octahedral tilting of the crystal. The $a^-b^+a^-$ mode (see Fig. 3(a)) composite three tilts: a '+' (in-phase) tilt about the b axis, and two '-' (out-of-phase) tilts around a and c axes by the same angle. Without tilting $a^0a^0a^0$ (see Fig. 3(d)), all octahedra motifs has the same orientation, the spin motif pair of the upper layer and the lower layer interconvert by both inversion and translation. Individual tilt $a^0b^+a^0$ around the b axis (see Fig. 3(b)) or the two tilts $a^-b^0a^-$ around a and c axes by the same angle (see Fig. 3(c)) also keeps the interconvertible relation. *The spin splitting in $LaMnO_3$ is, hence, a cooperative effect of the coexisting in-phase tilt(s) and out-of-phase tilt(s).*

We further studied how the magnitude of the tilting angles affects the resulting spin splitting in $LaMnO_3$. Fig. 3(e) shows the DFT calculated averaged spin splitting (averaged over all occupied bands on a dense k-mesh, see method for calculation details) for different structural configurations of in-phase and out-of-phase tilt angles. The left ($\alpha = \gamma = 0, \beta \neq 0$) and bottom ($\alpha = \gamma \neq 0, \beta = 0$) edge of the heatmap, corresponds to the decomposed tilt modes, show no spin splitting; while the middle of the heatmap ($\alpha = \gamma \neq 0, \beta \neq 0$), corresponds to the composite tilt modes, show finite spin splitting. The general trend of the heat map shows the spin splitting increases as the tilting angles become larger. The curved shape of the resulting spin splitting $\Delta_{SS}$ can be fitted to a product of the tilting angle about the a and c axes ($\alpha$ or $\gamma$) and the tilting angle about the b axis ($\beta$), that is $\Delta_{SS} = 0.32\alpha\beta$ (meV). We note that there is some small randomness in the heatmap that larger tilt angles could sometimes have smaller spin splitting, this might be attributed to the unavoidable octahedral distortions accompanied with tilting. Fitted equations like this will be very useful in predicting and designing perovskite materials having sizable spin splitting effect.

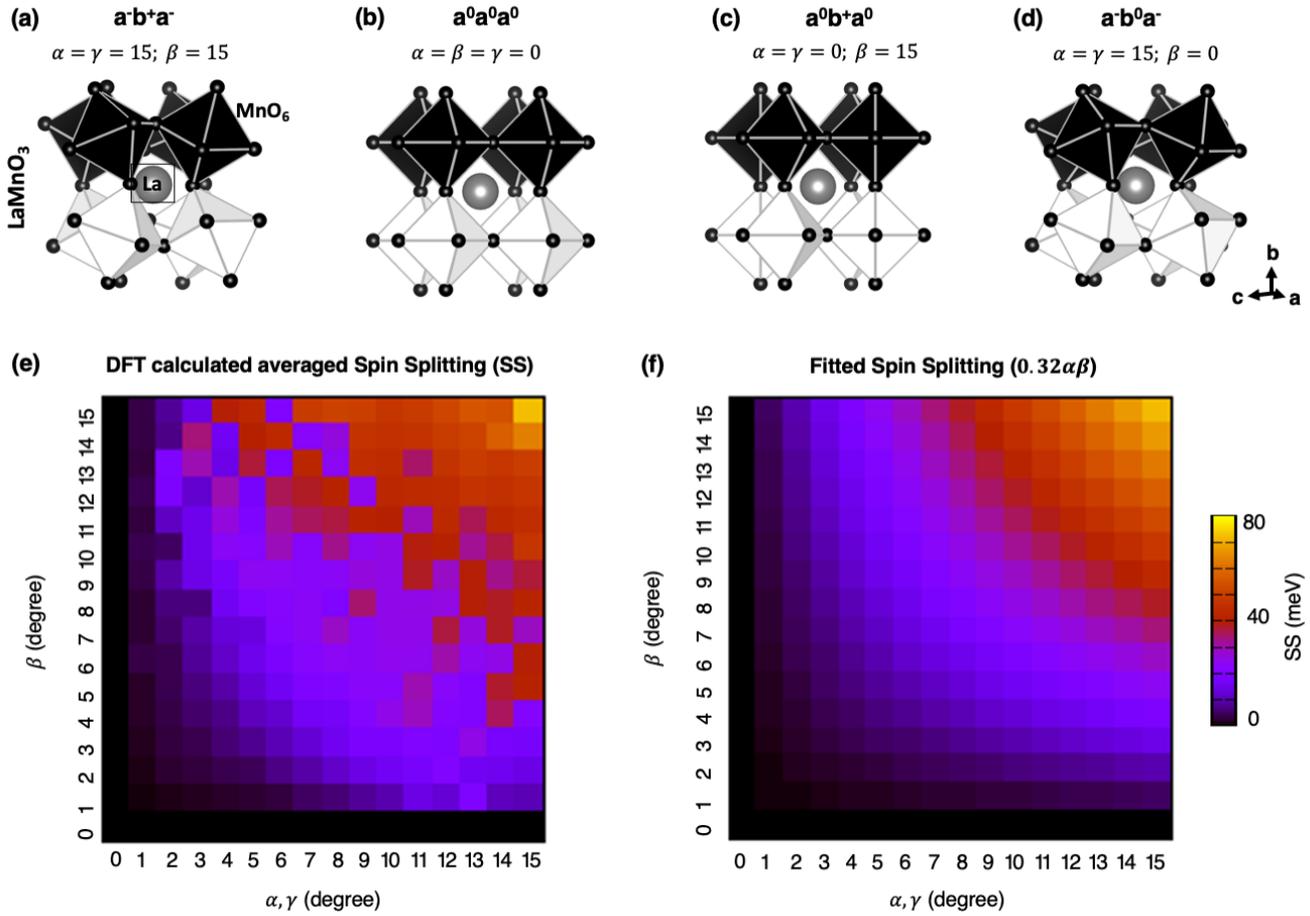

**Figure 3. Symmetry breaking induced by octahedral tilting in A-type antiferromagnetic orthorhombic LaMnO$_3$ crystal.** (**a**) Composite in-phase and out-of-phase tilt mode (a$^-$b$^+$a$^-$); (**b**) No tilt mode (a$^0$a$^0$a$^0$); (**c**) Individual in-phase tilt mode (a$^0$b$^+$a$^0$); (**d**) Individual out-of-phase tilt mode (a$^-$b$^0$a$^-$) model; (**e**) Density functional theory calculated averaged spin splitting (averaged over all occupied bands and k-points of a dense k-mesh) for different tilting angles of LaMnO$_3$; (**f**) Fitted spin splitting as product of tilting angles, that is $0.32\alpha\beta$.

Intriguingly, the spin motif pair also provides useful information about the momentum dependence of the spin splitting. This is accomplished by looking at the spin motif pair from different perspectives, i.e., projected spin motif pair. In fact, the spin splitting will only occur along wavevector directions where the shape of projected spin motif pair onto the corresponding planes (normal to the wavevector direction) are different. This reflects the correspondence between the reciprocal vector and the real space planes. For instance, in tetragonal NiF$_2$, as shown in Fig. 4, by projecting the two NiF$_6$ spin motif in the magnetic unit cell onto the different real space crystallographic planes, we see that the difference in shapes emerges when the spin motif pair are projected onto (110) and (1-10) planes but disappears when viewed from (100) or (010) or (001) perspective. This exactly maps to the k directions where the energy bands spin split (along [110] or [1-10] directions) and spin degenerate (along [100], [010], or [001]). Such correspondence between the "same vs different" projected spin motif pair shape and "degenerate vs split" energy bands thus provide a simple method to acquire information about at which momentum the spin splitting would occur, that otherwise requires sophisticated knowledges of the symmetries (beyond $\Theta I$ and $UT$) about the system [31] to figure out.

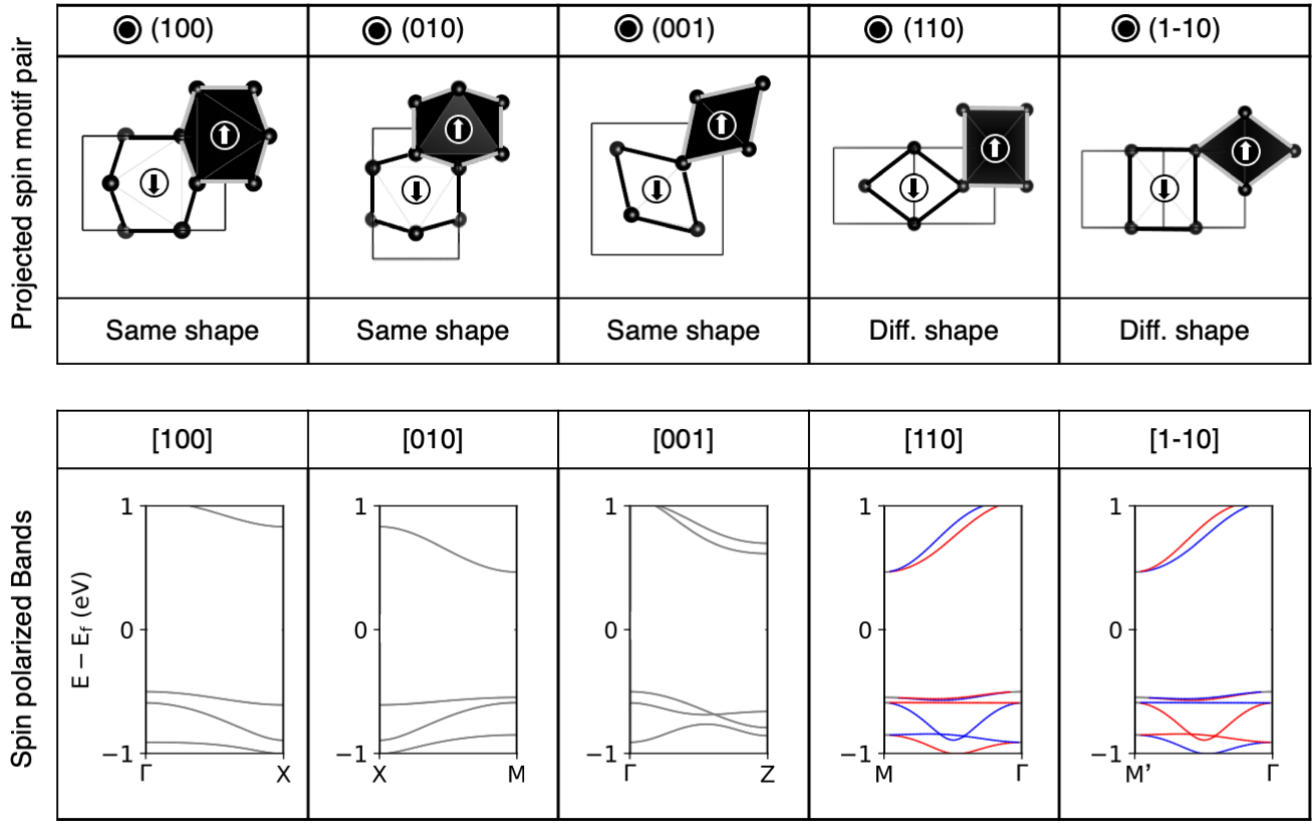

**Figure 4 | Projected spin motif pair of same vs different shape corresponds to momentum dependent spin degeneracy vs spin splitting.** In the graphics of projected spin motif pair in the upper panels, black and white shadings are used to distinguish the spin motif pair with opposite magnetic moments; Solid dots are used to represent F ions coordinate around Ni atoms. Spin polarized energy bands are shown in the lower panels where red and blue lines are used to map positively and negatively spin polarized bands. Grey lines are degenerate bands with no spin polarization.

## IV. Discussion

**A new form of magnetism?** It is interesting to ponder whether the occurrence of magnetic prototypes involving non-interconvertible spin motif pair qualify as a fundamentally new form of magnetic order relative to the known interconvertible spin motif pair. Recent work [31] advocated the idea of recognizing antiferromagnets with, what is defined here as, non-interconvertible spin motif pair as 'a third phase of magnetism' adding to the conventional antiferromagnetism and the ferromagnetism. Historically, antiferromagentism has been described by Louis Néel [32,33] as "antiparallel arrangement of atomic moments"; the concept was later generalized to describe a specific type of magnetism where "arrangement of atomic moments exactly compensates", including both collinear and noncollinear magnetic ordering. Following this traditional definition, it seems that the magnetic prototypes originating from interconvertible or non-interconvertible spin motif pair (listed in Fig A1, and [4,5]) belong to the same underlying antiferromagentism. Indeed, antiferromagnetic crystals come with many possible symmetries and different point group symmetries come with different materials properties. In this respect it does not seem consistent to distinguish systems with non-interconvertible geometry as a fundamentally different kind of magnetism with respect to antiferromagnetism. The traditional definition -- "arrangement of atomic moments exactly compensates" -- already encompasses both.

**Design of multifunctional materials:** The language of the motif emphasis on how atoms see each other in the local coordination environment, is closely associated with many material properties, therefore would facilitate the design of multifunctional materials.[23,24,34-38] For example, the tetragonal or octahedral coordination of $Mn^{2+}$ affects the formation of polarons that limits the electric conductivity [39]. The tetrahedral coordination associated with the $sp^3$ hybridization determines the electronic and optical properties of many zinc-blend semiconductors. [40,41]. The motif descriptor has also been applied in machine learning studies to predict electronic properties of materials [42]. Looking at the spin splitting effect from the perspective of spin motif pair not only offers a way to tell if an antiferromagnetic compound will enable the see whether symmetries are preserved or violated; but also helps to find future connection to other functionalities (such as bonding and stability) that are traditionally expressed by single motifs.

**Spin splitting in noncollinear antiferromagnets.** While the current paper focuses on the discussion of collinear antiferromagnetic compounds, we note that the spin splitting effect can also exist in noncollinear antiferromagnetic compounds. However, one should note that the symmetry condition of preserving *UT* in coplanar noncollinear antiferromagnets does not always guarantee spin degeneracy. Specifically, when the spin states are not aligned in the same plane of the coplanar plane, the *UT* symmetry may not reverse the spin states as it does in the collinear magnetic systems. This point needs to be considered when discussing about spin splitting effects in noncollinear antiferromagnets.

**Acknowledgements**
We thank Dr. Xiuwen Zhang for fruitful discussions. The work was supported by the National Science Foundation (NSF) DMR-CMMT Grant No. DMR-1724791 that supported the formal theory development of this work. The electronic structure calculations of this work were supported by the U.S. Department of Energy, Office of Science, Basic Energy Sciences, Materials Sciences and Engineering Division under Grant No. DE-SC0010467. This work used resources of the National Energy Research Scientific Computing Center, which is supported by the Office of Science of the U.S. Department of Energy.

**Methods**
**General Density function theory settings:** The electronic properties are calculated by the density function theory (DFT) method [43-45] implemented in the Vienna Ab initio Simulation package (VASP). For all the calculations, we employed the Perdew-Burke-Ernzerhof (PBE) exchange-correlation functional [46-48] with on-site coulomb on 3d orbitals following the simplified rotationally invariant approach introduced by Dudarev et al [49]. The atomic and magnetic structures of the four example compounds for the calculations are taken from X-ray scattering and/or neutron scattering experiments. The different configurations of tilted LaMnO3 models are constructed by cooperative tilt of the perfect cubic LaMnO$_3$ enforcing the bond length of Mn-O being 2 angstroms. Atomic positions and lattice vectors are fixed during the DFT self-consistent iterations. The magnetic configurations are simulated in collinear settings without turning on the spin-orbit coupling. We adopt a plane-wave basis of up to 500 eV energy cutoff, a $\Gamma$-centered k-mesh for hexagonal cell and Monkhorst-Pack [50] k-mesh otherwise for self-consistent charge density. We used the tetrahedron smearing method for insulators or semiconductors and Gaussian smearing method for metals.

**How we calculate the spin polarized density of states and band structures:** The density of states and energy bands are calculated non-self-consistently from a preconverged charge density. The density of states is evaluated for energy range that is a few eV above and below the fermi level, with NEDOS set to 2000. The energy bands are calculated on the conventional high-symmetry k-paths. The magnitude of the spin polarization of the bands (mapped by blue to red color) is calculated by projecting the eigenstates onto the direction of the common magnetic axis of the collinear antiferromagnetic compound.

**How we calculate the averaged spin splitting:** The strength of the spin splitting for one compound is evaluated as a weighted averaged in a self-consistent run. The sum of the spin splitting of the occupied bands for each k point is weighted by the multiplicity of the symmetry-reduced k points. The averaged spin splitting is then calculated as an average over all k points of the k sampling for all occupied bands, that is the sum of the spin splitting at different k divided by the number of bands and the number of k points.

**Appendix**

**A. From symmetries without SOC to symmetries with SOC**

Whereas traditionally people use magnetic space group (MSG) with SOC, the mathematical definition of MSG is not associated with the question of whether it's done with SOC or without SOC. We used both "MSG with SOC" and "MSG without SOC" (see Table II of Ref. [4]). The later refers to a subset of the spin space group symmetries consists of space-time operations or can be considered as the modified magnetic space group with pseudoscalar electron spin [51]. Although the MSG without SOC can be different from MSG with SOC, but the existence/absence of *UT* and *θI* is synchronized. That means it is valid to use the MSG with SOC to determine the spin splitting behavior of materials without SOC, this contrasts with the common sense that applying the magnetic symmetry with SOC to explain the SOC-independent spin splitting effect looks conceptually wrong. Moreover, in a collinear antiferromagnet the existence of *UT* in the zero SOC limit is equivalent to have a spatial translation *T* that reverses the spin arrangement of the compound but keeps the crystal structure invariant. Antiferromagnets with such primitive lattice translation are known as having black and white Bravais lattice and corresponds to MSG type-IV [13]. Therefore, the existence of *UT* in the zero SOC limit is equivalence to whether the MSG (with SOC) type is IV. We used these rules as filter in spin polarized band structure theory to illustrate unconventional momentum-dependent spin splitting in certain antiferromagnets [5], a spin splitting present even without spin-orbit coupling (SOC) or inversion symmetry breaking. Such approach creates a common language to distinguish traditional forms of spin splitting, as well as the unconventional "spin split" prototype. Similar classifications (despite some nuance difference) of spin degenerate vs spin splitting collinear magnets recently derived in Ref. [31] from spin symmetry based theory (where $UT$ is referred as [C2 | t]) for spin splitting collinear magnets without SOC also suggest the equivalence of the spin symmetry based approach and the magnetic space group based approach.

## B. Magnetic prototypes classified by the conditions of preserving or violating symmetries: The spin splitting types (SST)

Figure A1 illustrate the five magnetic prototypes of preserving/breaking of $\theta I$ and/or $UT$ symmetry and their correspondence to having interconvertible/non-interconvertible spin motif pair. The first prototype (SST-1) is represented by a pair of tetrahedra motifs (see Fig. A1(a)) that interconverts from one to the other by inversion. Because the tetrahedra are not inversion symmetric, the inversion related spin motif pair cannot be simultaneously interconverted by translation. This prototype corresponds to the symmetry condition of $UT$ violated and $\theta I$ preserved. The second prototype (SST-2) is given by a pair of octahedra motifs (see Fig. A1(b)) that also interconverts by inversion. Different from the tetrahedra, the octahedra motifs are inversion symmetric. It is easy to proof that the inversion symmetric spin motif pair that interconverts by inversion will also be interconverted by a translation. This holds for any centrosymmetric spin motif pair. The prototype corresponds to the symmetry condition of both $UT$ and $\theta I$ preserved. This prototype has been illustrated in Fig. 1(a). The third prototype (SST-3) is illustrated by a pair of tetrahedra motifs (see Fig. A1(c)) that interconverts by translation. Again, because the tetrahedra are not inversion symmetric, the translation related spin motif pair cannot be simultaneously interconverted by inversion. This prototype corresponds to the symmetry condition of $UT$ preserved and $\theta I$ violated. The fourth prototype (SST-4) is the most interesting prototype where SOC independent spin splitting occurs. The prototype is represented by a pair of octahedra motifs (see Fig. A1(d)) that neither connected by translation nor inversion. This prototype has been illustrated in Fig. 1(b). We note, for this prototype, there is not restriction on the shape of the motif being centrosymmetric or non-centrosymmetric. This prototype corresponds to the symmetry condition of both $UT$ and $\theta I$ violated. The fifth prototype (SST-5) is the ferromagnetic prototype where Zeeman-type spin splitting occurs. The prototype is represented by two octahedra motifs (see Fig. A1(e)) of the same spin (color) that neither connected by translation nor inversion. This prototype has been illustrated in Fig. 1(c).

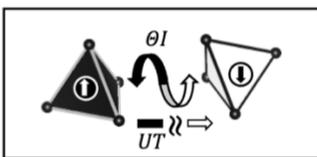

| Proto-type | Magnetism | Has $UT$ | Has $\theta I$ | Spin splitting | Illustration |
|---|---|---|---|---|---|
| SST-1 | AFM | X | ✓ | No SS | (a) |
| SST-2 | AFM | ✓ | ✓ | No SS | (b) |
| SST-3 | AFM | ✓ | X | No SS | (c) |
| SST-4 | AFM | X | X | SS | (d) |
| SST-5 | FM | X | X | SS | (e) |

Centrosymmetric (CS) and Non-CS spin motif

$\theta$: Time reversal   $I$: Spatial inversion
$U$: Spin inversion   $T$: Non-primitive translation

**Figure A1| Schematic illustration of magnetic prototypes preserves/violates $\theta I$ and/or $UT$ symmetry and having interconvertible or non-interconvertible spin motif pair. (**a) antiferromagnets with spin motif pair interconverts by inversion but not by translation; (b) antiferromagnets with spin motif pair interconverts by inversion and translation; (c) antiferromagnets with spin motif pair interconverts by translation but not by inversion; (d) antiferromagnets with spin motif pair fails to interconvert by neither inversion nor translation. (e) ferromagnets with spin motif pair fails to interconvert by neither inversion nor translation. The octahedra and tetrahedra within the magnetic unit cell are used to represent centrosymmetric and non-centrosymmetric spin motifs; the arrow inside each motif is used to indicate the magnetic moment orientation associated with the motif. The geometric interconvertible relation between spin motif pair is indicated by line arrow (translation related) and S-shaped arrows (inversion related) between them. Broken arrows mean the interconvertible relation does not exist.

## C. DFT calculated spin polarized band structure for the four real antiferromagnetic compounds

Figure A2 shows the spin polarized energy bands calculated using PBE+U method for the four antiferromagnetic compounds used as examples in the main text. Fig. A2(a-c) shows for the three compounds CuMnAs, FeBr$_2$, and BiCoO$_3$ consists of interconvertible motif pair their energy bands are spin degenerate and have no spin polarization. Fig. A2(d) shows for NiF$_2$ consists of non-interconvertible motif pair its energy bands are spin split and has finite spin polarization. These results agree well with the symmetry and motif description.

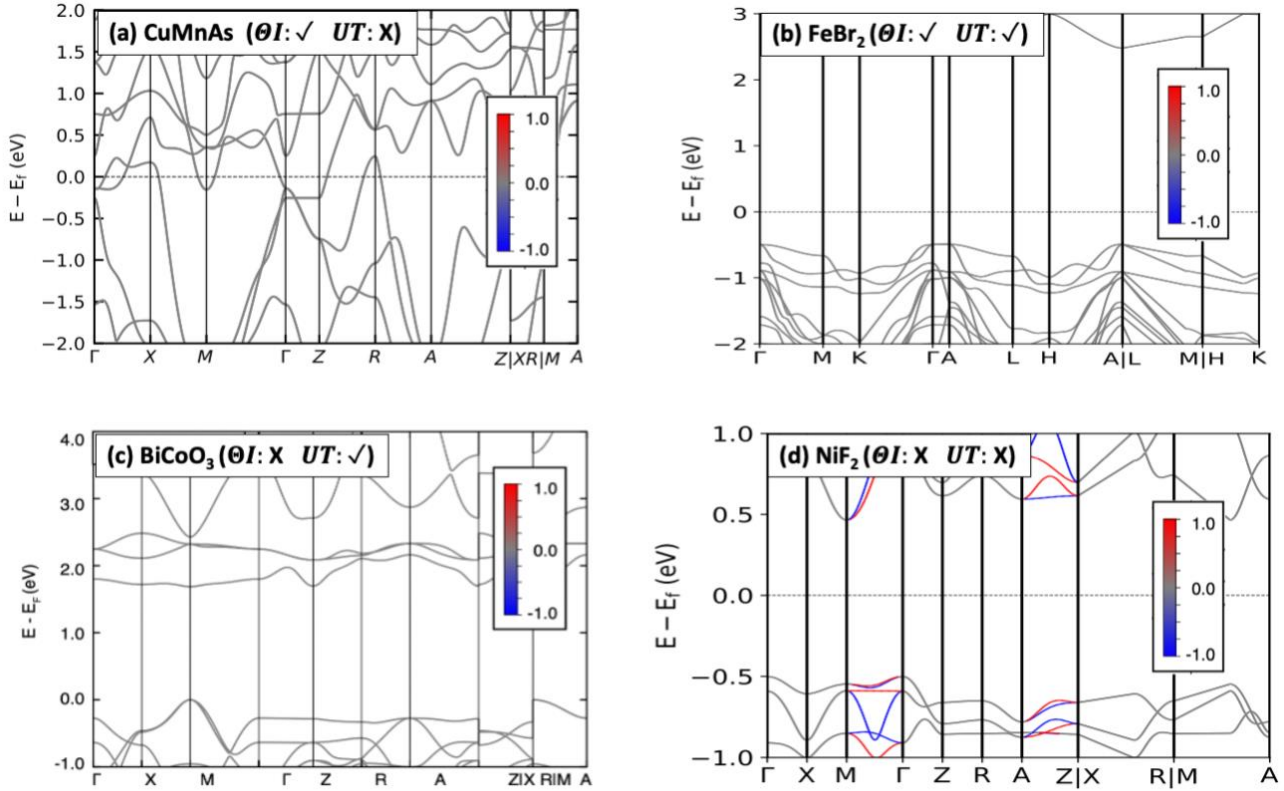

**Figure A2 | DFT calculated spin polarized band structure with SOC turned off for four examples of different cases.** Antiferromagnetic compounds (a) CuMnAs, (b) FeBr$_2$, and (c) BiCoO$_3$ with "interconvertible" spin motif pair does not show spin splitting without SOC; (d) NiF2 with "non-interconvertible" spin motif pair shows spin splitting without SOC. The spin polarization bands are mapped from blue (negatively spin polarized) to grey (zero spin polarized) to red (positively spin polarized).